%% file: main.tex
\documentclass{article} %
\usepackage{iclr2024_conference}
\usepackage{times}

\input{math_commands.tex}

\usepackage[hidelinks]{hyperref}
\usepackage{url}
\usepackage{soul,color}

\usepackage{float}
\usepackage{lipsum}
\usepackage{todonotes}
\usepackage{booktabs}
\usepackage{multirow}
\usepackage{makecell}
\usepackage[ruled,vlined]{algorithm2e}
\usepackage{wrapfig}
\usepackage{amsthm}
\usepackage{tikz}
\usepackage{subcaption}
\usepackage{colortbl}
\usetikzlibrary{fit,calc}

\title{Physical Encoding Improves OOD Performance in Deep Learning materials property prediction} %

\iclrfinalcopy
\author{%
  Nihang Fu, Sadman Sadeed Omee, Jianjun Hu\\
  Department of Computer Science and Engineering, University of South Carolina, Columbia, SC, USA\\
  \texttt{Correspondence: jianjunh@cse.sc.edu}
}

\begin{document}

\maketitle

\begin{abstract}

Deep learning (DL) models have been widely used in materials property prediction with great success, especially for properties with large datasets. However, the out-of-distribution (OOD) performance of such models are questionable, especially when the training set is not large enough. Here we showed that using physical encoding rather than the widely used one-hot encoding can significantly improve the OOD performance by increasing models' generalization performance, which is especially true for models trained with small datasets. Our benchmark results of both composition- and structure-based deep learning models over six datasets including formation energy, band gap, refractive index, and elastic properties predictions, demonstrated the importance of physical encoding to OOD generalization for models trained on small datasets.
\end{abstract}

\section{Introduction}

Composition-based deep learning models, such as Roost \cite{goodall2020predicting} and CrabNet \cite{wang2021compositionally, Wang2022explainablegap}, and structure-based deep learning models, such as CGCNN \cite{xie2018crystal}, ALIGNN \cite{choudhary2021atomistic}, DeeperGATGNN \cite{omee2022scalable}, and coGN \cite{ruff2024connectivity}, have been widely used in materials property prediction. A benchmark leaderboard, Matbench \cite{dunn2020benchmarking}, has been set up to show the performances of different models over a list of composition- and structure-based materials property prediction problems. Here, atomic information serves as an essential feature for training all of these models. Diverse encoding methods are employed to encode atomic information, such as one-hot encoding, Magpie encoding, Matscholar encoding, and feature encoding including different atom properties, etc. For example, in Roost, the default encoding method employed Matscholar embedding to represent each node in the atomic graph \cite{goodall2020predicting}; in DeeperGATGNN, one-hot encoding set encoding different atoms by setting specific bit as the node feature of the input \cite{omee2022scalable}; CGCNN adopted a comprehensive approach, wherein each node was represented by a feature vector containing 9 atomic properties, including group number, period number, electronegativity, covalent radius, valence electrons, first ionization energy, electron affinity, block, and atomic volume \cite{xie2018crystal}; ALIGNN got inspiration from CGCNN's encoding method, employing the same atomic encoding with 9 input node features derived from atomic species \cite{choudhary2021atomistic}; coGN integrated the atomic number alongside atom features, such as atomic mass, radius, electronegativity, ionization, and oxidation states, and then they found that the prediction accuracy increases when they include more atom features in addition to the atomic number \cite{ruff2024connectivity}. In these publications, the authors usually reported that the one-hot-like encoding actually can give them the best performance along with their simplicity. However, one-hot-like encoding usually encodes atomic information to a sparse vector, and this sparsity prompts us to question whether one-hot-like encoding is the optimal encoding method for achieving strong prediction performance, especially for the performance of models trained on small datasets and evaluated on OOD samples.

Out-of-distribution (OOD) data is crucial in machine learning since it is pivotal for evaluating models' capabilities in generalization, robustness, and reliability. In many areas, where the aspiration is to build models that transcend the confines of training data and perform adeptly in real-world scenarios, the evaluation of OOD data emerges as a paramount challenge. It serves as a litmus for the generalization of a model to new contexts and helps to test its ability to extrapolate beyond familiar data distribution. There have been numerous papers to conclude the utilization of OOD set across various applications of machine learning in diverse domains. For example, Gui et al. \cite{koh2021wilds} developed a graph OOD benchmark to show a big gap between in-distribution and OOD graph data, such as molecules, sentence parser trees, webpage networks, and so on. Koh et al. \cite{koh2021wilds} discussed a broader application scope for OOD data beyond graph data. Introducing a benchmark named WILD, they aimed to capture a diverse array of distribution shifts inherent in real-world scenarios. Examples include shifts across hospitals for tumor identification and across time and location in satellite imaging and poverty mapping. Akin to the other fields, in materials science, the quest often extends beyond the known materials. Unknown new materials in material space may present compositions, structures, or properties vastly distinct from those encountered in existing materials \cite{omee2024structure}. Consequently, the ability of a model to handle OOD data becomes indispensable, since it allows researchers to predict and navigate novel materials. Recently, some researchers have worked on OOD prediction problems in materials science. Li et al. \cite{li2023critical} found that ML models trained on Material Project 2018 cannot perform well on new materials in Material Project 2021 because of the distribution shift. Hu et al. \cite{hu2024realistic} employed domain adaptation (DA) to improve the performance of existing ML models for materials property prediction and they evaluated trained models on a set of five OOD test set that they created. Their experiments demonstrate that employing DA can substantially enhance the prediction performance of ML models on OOD test sets. Omee et al. \cite{omee2024structure} proposed a benchmark showing that current state-of-the-art graph neural network (GNN) exhibit performance degradation on the OOD property prediction tasks compared to their baselines in the MatBench study, demonstrating a crucial generalization gap in realistic material prediction tasks. They also examine the latent physical spaces of these GNN models and analyze the reason behind the different performances. Shimakawa et al. \cite{shimakawa2024extrapolative} proposed a benchmark revealing that conventional ML models exhibit remarkable performance degradation beyond the training distribution, especially for small-data properties. To address this, they introduced a quantum-mechanical (QM) descriptor dataset (QMex) and an interactive linear regression (ILR) to incorporate interaction terms between QM descriptors and categorical molecular structures information, achieving state-of-the-art extrapolative performance while maintaining interpretability. In this paper, we take a novel perspective to analyze OOD data by proposing new OOD selection methods for both composition- and structure-based models. Additionally, unlike previous studies, we aim to demonstrate that input encoded with different encoding methods, particularly encoding with physical information, enables models to better handle OOD data.

To sum up, our contributions in this paper are summarized as follows:

\begin{itemize}
    \item We propose new OOD sample selection methods for composition- and structure-based models, respectively, and discuss how different OOD selection methods affect the model training.
    \item We argue that the one-hot element encoding or one-hot-like encoding is actually not the ideal configuration for achieving strong prediction performance for OOD samples, especially in the case of small training datasets. 
    \item We observe that models trained with different encoding methods exhibit varying performances on different types of OOD datasets, which may differ in property range or elemental distribution. This observation helps us to select encoding methods tailored to the specific type of OOD data, thereby enhancing predictions.
\end{itemize}

\section{Method}
This section outlines the methods used to generate the datasets, the four different encoding methods for atomic information, and the two networks, Roost (a composition-based network) and ALIGNN (a structure-based network), employed in this paper.

\begin{table}[!htb]
\caption{Details of datasets used in this work.}
\label{tab:datasets}
\begin{center}
\begin{small}
\begin{tabular}{cccc}
\toprule
        & Property (Unit)  & Source     & Total Samples  \\ \midrule
Matminer Dielectric   & Band Gap (eV) & Matminer  \cite{ward2018matminer, petousis2017high}          & 1,056     \\
Matminer Elasticityity  & Elastic Anisotropy (-)  & Matminer \cite{ward2018matminer, de2015charting}           & 1,181            \\

\midrule

Dielectric  & Refractive Index (-)  & Matbench \cite{dunn2020benchmarking, petousis2017high}           & 4,764          \\
Perovskites  & Formation Energy (eV/atom) & Matbench \cite{dunn2020benchmarking, castelli2012new}             & 18,928          \\
GVRH    & Shear Modulus (GPa) & Matbench \cite{dunn2020benchmarking, de2015charting}             & 10,987             \\
KVRH   & Bulk Modulus (GPa)  & Matbench \cite{dunn2020benchmarking, de2015charting}            & 10,987         \\ 
\bottomrule
\end{tabular}
\end{small}
\end{center}
\end{table}

\subsection{Datasets}
\label{subsec:datasets}

\begin{figure}[!htb] 
    \centering
        \includegraphics[width=0.85\textwidth]{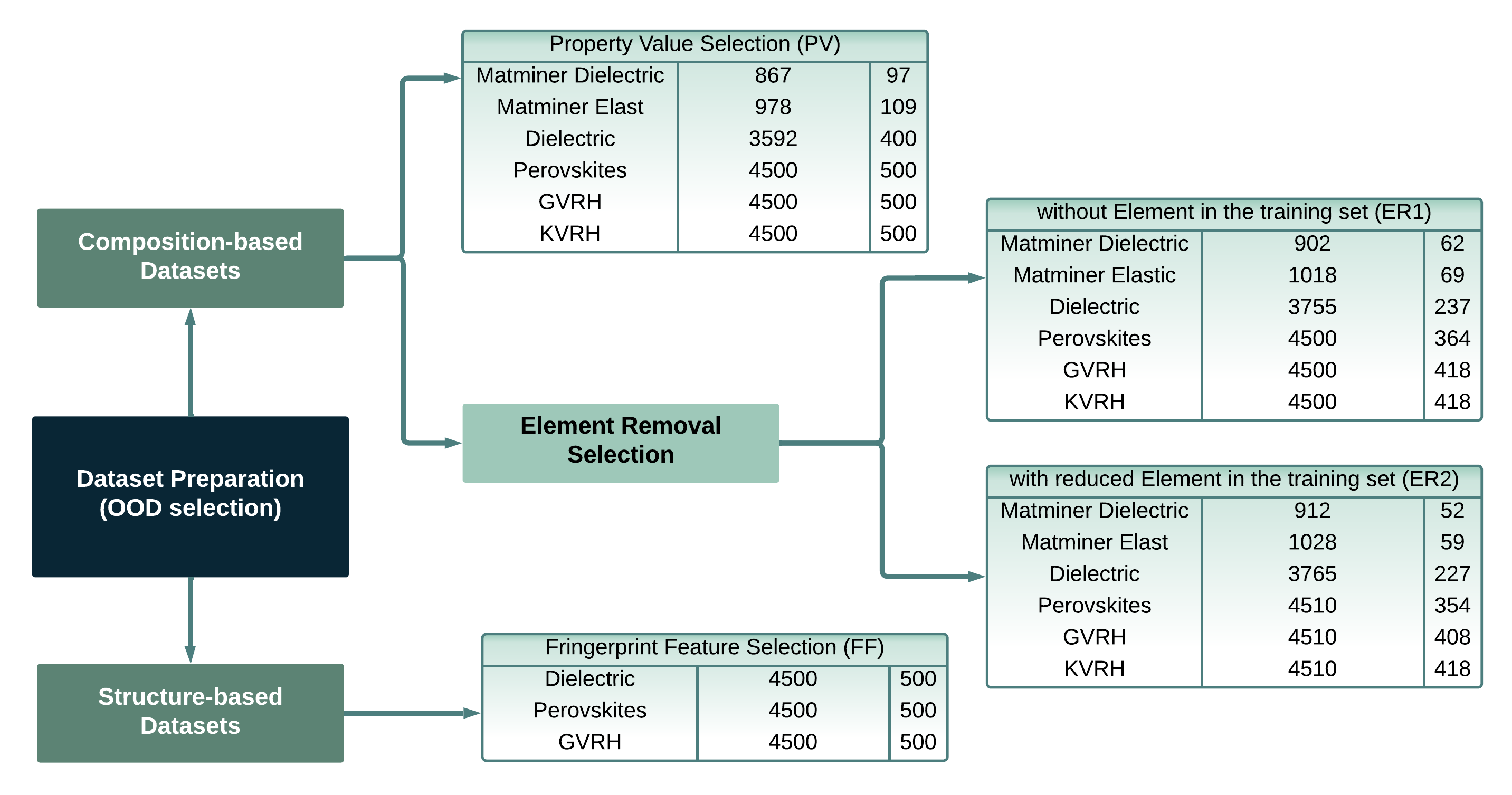}
     \caption{An overview of dataset preparation methods. Depending on the chosen models, we prepared composition-based and structure-based datasets. For each, we use different OOD selection methods (see details in Section \ref{subsec:datasets}). The size of each dataset for PV, ER1, ER2, and FF selection methods are shown in the figure. The first column lists the dataset names; the second column indicates the number of materials in the training set; the third column specifies the number of materials in the OOD set.}
     \label{fig:pipeline}
\end{figure}

We selected six datasets from Matminer \cite{ward2018matminer} and Matbench \cite{dunn2020benchmarking} as shown in Table \ref{tab:datasets} and prepared composition- and structure-based datasets (as shown in Fig \ref{fig:pipeline}) from these datasets to conduct experiments for training a composition-based network (Roost) and a structure-based network (ALIGNN) using four different encoding methods (See Section \ref{subsec:encoding} for details) for the inputs.

\paragraph{Composition-based datasets} 

To ensure a well-defined composition-to-property mapping, we first include only one polymorph per composition in each dataset. Next, we select OOD test sets based on two primary criteria: property value and element removal (the sizes of datasets as shown in the top and middle parts in Fig \ref{fig:pipeline}, respectively). These factors guide the selection of OOD test sets. We then selectively choose small subsets from the remaining data to train the models.

\begin{itemize}
    \item \textbf{Property value (PV)} The PV method involves sorting materials in each dataset according to their property values. Since there are significant differences exist between materials with varying properties, we can effectively meet our out-of-distribution requirements by using materials with higher property values as the OOD set. For datasets containing 5000 or more samples, we select the first 4500 data for model training and the last 500 data as the OOD set. For smaller datasets with fewer than 5000 samples, we use the first 90\% of the data for training and the remaining 10\% as OOD data. This approach ensures a significant difference in the distribution of property values between the training data and the OOD data.
    \item \textbf{Element removal (ER)}.  TheER method is to either remove compositions containing an element or reduce the percentage of this element within the datasets to compose OOD sets. The basic idea behind this selection method is that the model trained on a dataset With reduced or no representation of a specific element becomes less familiar with this elemental information, making it challenging to predict properties for compositions containing that element. This allows us to test the stability and generalization of the trained models using different encoding methods across these OOD sets. We begin by removing compositions containing a specific element (e.g., Ca in our experiment) to construct the ER1 OOD test set. For the ER2 OOD set, we only reduce the percentage of compositions containing Ca. Then, we randomly select small subsets from the remaining data to explore the performance of models trained on small datasets. We conduct five rounds of training, where for each round, we randomly split the selected subsets into a training set and an in-distribution (ID) test set in a 9:1 ratio. After training, we evaluate the five trained models on both their ID test sets and the corresponding OOD test sets to compare their average performance.
\end{itemize}

\paragraph{Structure-based datasets} 
For the purpose of exploring the performance of models trained on small datasets, we randomly select subsets from the original datasets (the sizes of datasets as shown in the bottom part of Fig \ref{fig:pipeline}). We calculate fingerprint features for materials in each dataset and determine the distance between material pairs based on these fingerprint features. By analyzing pair distances, we identify the three closest neighbors for each material. Subsequently, we calculate the average distance to these three neighbors. To build OOD test sets, we want to find the sparsest samples, so we sort these average neighbor distances and select 500 samples with the highest average distance from neighbors. Then we randomly select 4500 samples from the remaining data to train the Roost model. During the training process, we randomly split each dataset into a training set and an ID test set at a 9:1 ratio and repeated the training process five times. Subsequently, we evaluate the five trained models on their ID test sets and respective OOD test sets to compare their average performance.

\subsection{Element encoding methods}
\label{subsec:encoding}

In this work, we consider four different encoding methods to represent atomic information in material graphs.

\begin{itemize}
    \item \textbf{One-hot encoding.} One-hot encoding is a widely used method in graph neural networks (GNNs) for material property prediction. In this encoding method, each node/atom in the graph is encoded as a binary vector where only one position in this binary vector is set to 1, while all other positions are set to 0.  For example, in one-hot encoding, a binary vector with a length of 100 is generated to represent one atom. We consider the element Hydrogen ($H$) with only the first position of 1, while all other positions should be 0. This encoding method does not contain any physical information about the material, only using different bits in the binary vector to represent different atoms. Additionally, The generated binary vectors are usually too sparse, which may lead to suboptimal results.
    \item \textbf{CGCNN encoding.} CGCNN encoding method is a physical atomic encoding method from the CGCNN paper \cite{xie2018crystal}, which concatenates multiple one-hot encoding vectors for 9 different atomic properties: group number, period number, electronegativity, covalent radius, valence electrons, first ionization energy, electron affinity, block, and atomic volume. In detail, the group number spans 18 categories from 1 to 18, thus it can be represented as a binary vector with a length of 18. Similarly, the period number has 9 categories ranging from 1 to 9; electronegativity is categorized into 10 categories ranging from 0.5 to 4.0; covalent radius has 10 categories ranging from 25 to 250; valence electrons are classified into 12 categories from 1 to 12; first ionization energy has 10 categories with a range from 1.3 to 3.3; electron affinity has 10 categories with a range from -3 to 3.7; block has 4 categories that is p, d, and f; atomic volume is segmented into 10 categories with a range from 1.5 to 4.3 \cite{xie2018crystal}. Combining binary vectors of these properties, we obtain the CGCNN encoding, which comprises a vector of length 93. This encoding method allows GNN models to capture more atomic information in addition to the atomic composition information of the material. 
    \item \textbf{Matscholar encoding.} Matscholar encoding is a natural language processing (NLP) embedding based on the data mining of definite compositions and structure prototypes \cite{tshitoyan2019unsupervised, weston2019named}. A large corpus of material science-related documents is collected, and a machine learning model is trained on the preprocessed text. This model learns to predict the context of a word based on its surrounding words, capturing a wide range of material knowledge, including material composition, properties, processing methods, and the relationships between these different pieces of knowledge. Unlike the traditional one-hot encoding method, which uses a sparse binary vector, Matscholar embedding utilizes a dense vector representation for each atom. This encoding represents various atomic knowledge in a continuous vector of length 200, allowing for efficient representation of atomic information. This encoding has a higher sense of physical encoding than the one-hot encoding but is not as direct as the CGCNN encoding.
    \item \textbf{MEGNet encoding.}  This is a neural encoding method derived from MEGNet, a structure-based GNN model for property prediction of both molecules and crystals \cite{chen2019graph}. The original input vector of the MEGNet model for crystal property prediction utilized classical one-hot encoding for elements with a length of 94. The MEGNet embedding is generated from the embedding layer in the MEGNet model trained on a task to predict formation energies of about 69,000 materials from the Materials Project (MP) Database \cite{goodall2020predicting, chen2019graph}. The generated MEGNet encoding vector contains the element features for 89 elements and its length for each element is 16. 
    
\end{itemize}

\subsection{Deep learning models for materials property prediction}
We utilized one composition-based deep learning model (Roost) and one structure-based deep learning model (ALIGNN) to evaluate the performance of models using different encoding methods.

\paragraph{Roost} Representation Learning from Stoichiometry (Roost) \cite{goodall2020predicting} is a GNN-based network that takes only the composition as input. Each material will be represented as a stoichiometry graph as the input of the Roost network. Each element in the model's input domain is represented by a vector. The final entry in the initial internal representation is the fractional weight of the element. A weighted soft-attention mechanism is then used to update these elemental representations.

\paragraph{ALIGNN} %

Atomistic Line Graph Neural Network (ALIGNN) \cite{choudhary2021atomistic} is a GNN-based network for structure based material property prediction, which performs message passing on both the interatomic bond graph and its line graph corresponding to bond angles. The input of ALIGNN is to represent each material by a graph with node features, edge features, and triplet features. ALIGNN alternates between graph convolution on graphs and line graphs to propagate bond angle information through interatomic bond representations to the atom-wise representations and vice versa. We use both the bond distances and angles in the line graph to incorporate finer details of atomic structure which leads to higher model performance.

\subsection{Evaluate metrics}
We utilize the Mean Absolute Error (MAE), Root Mean Square Error (RMSE), and R-squared ($R^2$) scores to evaluate the performance of each model.

\begin{equation}
MAE = \frac{\Sigma_{n=1}^N |\hat{y_n} - y_n|}{N}
\end{equation}

\begin{equation}
RMSE = \sqrt{\frac{\Sigma_{n=1}^N (\hat{y_n} - y_n)^2}{N}}
\end{equation}

\begin{equation}
R^2 = 1 - \frac{\sum_{n=1}^{N} (y_n - \hat{y}_n)^2}{\sum_{n=1}^{N} (y_n - \bar{y})^2}
\end{equation}

where $N$ is the the total number of data points; $\hat{y_n}$ and $y_n$ denote the prediction and target values of the $n^{th}$ material, respectively; $\bar{y}$ is the mean values of all target values. 

\vspace{-6pt}  
\section{Results}\label{sec:experiments}
\vspace{-6pt}

In this section, we present and analyze the performance of Roost and ALIGNN based on the Mean Absolute Error (MAE), Root Mean Square Error (RMSE), and R-squared ($R^2$) scores. The quantitative and qualitative results for various datasets and models are summarized below.  %

\subsection{Performance comparison of models for composition-based materials property prediction}

Table \ref{tab:pv}, Table \ref{tab:er1}, and Table \ref{tab:er2} compare the prediction performance of Roost trained on different datasets selected using PV, ER1, and ER2 OOD selection methods with different atomic encoding methods, respectively. By comparison, we can see that the composition-based models trained using encoding methods that include physical information (e.g., CGCNN, Matscholar, and MEGNet) usually perform better and are more stable than those trained with the one-hot encoding method. We also find that models trained and tested on in-distribution data exhibit varying performances across different OOD selection methods, especially for models trained with the one-hot encoding methods.

\begin{table}[!htb]
\caption{Results of composition-based datasets: PV selection method.}
\label{tab:pv}
\begin{center}
\begin{small}
\begin{tabular}{ccrrrrrr}
\toprule
&   &  \multicolumn{2}{c}{MAE $\downarrow$}                                & \multicolumn{2}{c}{RMSE 	$\downarrow$}                  & \multicolumn{2}{c}{$R^2$ Score 	$\uparrow$} \\ 
 \cmidrule(lr){3-4} \cmidrule(lr){5-6} \cmidrule(lr){7-8}
&  & \multicolumn{1}{c}{ID} & \multicolumn{1}{c}{OOD} & \multicolumn{1}{c}{ID} & \multicolumn{1}{c}{OOD} & \multicolumn{1}{c}{ID} & OOD     \\ 
\midrule
 & One-hot    & 0.5621                 & 2.4095                 & 0.7715                 & 2.6311                  & 0.5320                    & -5.7439                   \\ 
 & CGCNN      & 0.5183                 & \textbf{2.1886}        & 0.7139                 & 2.5064                  & 0.5986                   & -5.1393                   \\
 & Matscholar & \textbf{0.4895}        & 2.2322                 & 0.6887                 & \textbf{2.4836}         & 0.6267                   & \textbf{-5.0361}          \\
\multirow{-4}{*}{\cellcolor[HTML]{FFFFFF}Matminer Dielectric}                                     & MEGNet     & 0.4996                 & 2.3434                 & \textbf{0.6799}        & 2.6443                  & \textbf{0.6358}          & -5.8079                   \\
\midrule

& One-hot    & 0.4149                 & 10.1634                & 0.6132                 & 43.7523                 & -0.6974                  & -0.0601                   \\
& CGCNN      & 0.4069                 & \textbf{10.0876}       & 0.6050                  & 43.7346                 & -0.6511                  & -0.0593     \\
 & Matscholar & 0.3865                 & 10.1372                & 0.5787                 & 43.7546                 & -0.5113                  & -0.0602                   \\
\multirow{-4}{*}{\cellcolor[HTML]{FFFFFF}Matminer Elasticity}                                   & MEGNet     & \textbf{0.3742}        & 10.1474                & \textbf{0.5539}        & \textbf{43.7171}        & \textbf{-0.3826}         & \textbf{-0.0584}          \\
\midrule

 &  One-hot    & 0.1611                 & \textbf{3.2232}        & 0.2790                  & \textbf{6.8455}         & 0.7199                   & \textbf{-0.3369}          \\
 & CGCNN      & \textbf{0.1332}        & 3.5505                 & \textbf{0.2115}        & 7.0351                  & \textbf{0.8394}          & -0.4121                   \\
 & Matscholar & 0.1355                 & 3.4872                 & 0.2287                 & 6.9943                  & 0.8124                   & -0.3957                   \\
\multirow{-4}{*}{\cellcolor[HTML]{FFFFFF}Dielectric}          & MEGNet     & 0.1349                 & 3.4909                 & 0.2143                 & 6.9922                  & 0.8350                    & -0.3948 \\
\midrule
& One-hot    & 0.0625                 & 1.1908                 & 0.0912                 & 1.2411                  & 0.8750                    & -10.0855   \\
 & CGCNN      & 0.0565                 & \textbf{1.1051}        & 0.0832                 & \textbf{1.1489}         & 0.8959                   & \textbf{-8.5121}          \\
 & Matscholar & \textbf{0.0549}        & 1.2492                 & \textbf{0.0806}        & 1.2960                  & \textbf{0.9022}          & -11.1275                  \\
\multirow{-4}{*}{\cellcolor[HTML]{FFFFFF}Perovskites}     & MEGNet     & 0.0582                 & 1.1564                 & 0.0849                 & 1.2055                  & 0.8914                   & -9.4518                   \\
\midrule
& One-hot    & 0.1323                 & 0.8124                 & 0.1831                 & 0.8495                  & 0.5741                   & -94.7312                  \\ 
 & CGCNN      & 0.1181                 & \textbf{0.7550}        & 0.1692                 & \textbf{0.7751}         & 0.6363                   & \textbf{-78.6911}         \\
 & Matscholar & 0.1178                 & 0.7724                 & 0.1689                 & 0.8002                  & 0.6373                   & -83.9460                  \\
\multirow{-4}{*}{\cellcolor[HTML]{FFFFFF}GVRH}     & MEGNet     & \textbf{0.1141}        & 0.7839                 & \textbf{0.1682}        & 0.8029                  & \textbf{0.6404}          & -84.5450                  \\
\midrule
& One-hot    & 0.1118                 & 0.6961                 & 0.1859                 & 0.7370                  & 0.5698                   & -143.3415                 \\ 
 & CGCNN      & 0.1051                 & \textbf{0.6855}        & \textbf{0.1778}        & \textbf{0.7099}         & 0.6066          & \textbf{-132.8270}        \\
 & Matscholar & 0.1076                 & 0.7173                 & 0.1846                 & 0.7396                  & 0.5760                    & -144.4056                 \\
\multirow{-4}{*}{\cellcolor[HTML]{FFFFFF}KVRH}     & MEGNet     & \textbf{0.1027}        & 0.7312                 & \textbf{0.1778}        & 0.7476                  & \textbf{0.6064}          & -147.6026                 \\

\bottomrule
\end{tabular}
\end{small}
\end{center}
\end{table}

In Table \ref{tab:pv}, we present the results of Roost trained on datasets selected using the PV method. For all metrics on the in-distribution (ID) test sets, models with one-hot encoding underperform compared to the other three encoding methods on most datasets. For example, on the Matminer Dielectric dataset, the MAE, RMSE, and $R^2$ score of the model with one-hot encoding are 0.5621, 0.7715, and 0.5320, respectively, which are 8.45\%, 8.07\%, and 11.13\% worse than the model with the CGCNN encoding, the encoding method with the closest performance, and 12.51\%, 13.47\%, and 16.33\% worse than the model with the Matscholar encoding, the best-performing encoding method. However, for OOD test sets, the performance of each model deteriorates significantly. We observe that models trained on Matminer Dielectric, Matminer Elasticity, Dielectric, and Perovskites datasets exhibit MAE and RMSE values larger than 1. Especially, on the Matminer Elasticity dataset, the MAE and RMSE exceed 10 and 43, indicating that these models essentially fail to capture features of OOD data selected by PV methods. On the GVRH and KVRH datasets, despite the MAE and RMSE being comparable, the $R^2$ score can degraded to around -100, further reflecting the poor performance of these trained models on the OOD test sets selected by the PV method. This decline makes sense, as the PV selection method chooses OOD samples with target values different substantially from those in the training set, making it challenging for the composition-based model to predict such different values accurately.

Table \ref{tab:er1} shows the results of Roost trained on datasets selected using the ER1 method. On the Matminer Dielectric dataset, the performance differences among models using the three physical encoding methods are within 5.56\% on the in-distribution (ID) set, but the one-hot encoding model shows MAE, RMSE, $R^2$ score of 0.5905 eV, 0.8498 eV, and 0.7055 eV, which are worse than the model using Matscolar encoding (the closest in performance) by 13.93\%, 10.16\%, and 6.23\%, respectively, let alone the best performing-ones. On the OOD set of the Matminer Dielectric dataset, the model using one-hot encoding performs the worst, while the model using MEGNet encoding performs the best. Models using one-hot encoding exhibit significant performance degradation on the OOD set, with MAE, RMSE, and $R^2$ score declining by 39.61\%, 31.11\%, and 34.35\% from the ID set. In contrast, the model using CGCNN declines by 12.72\%, 4.63\%, and 3.19\%, respectively; the model using Matscholar encoding declines by 8.7\%, 4.63\%, and 3.74\%, respectively; the model using MEGNet encoding declines by 10.58\%, 2.14\%, and -0.23\%, respectively. Similar trends are observed in the Perovskites, GVRH, and KVRH datasets. However, for the Matminer Elasticity and Dielectric datasets, the $R^2$ scores for both ID and OOD sets indicate that these models fail to capture elastic anisotropy and refractive index related information, resulting in poor performance on both sets. Overall, models using physical encoding methods are more stable than those using one-hot encoding with the ER1 selection method. Among the physical encoding methods, models with CGCNN encoding (the one-hot-like encoding) are less stable than models with Matscholar encoding, and models with Matscholar encoding are less stable than models using MEGNet encoding.

\begin{table}[!htb]
\caption{Results of composition-based datasets: ER1 selection method.}
\label{tab:er1}
\begin{center}
\begin{small}
\begin{tabular}{ccrrrrrr}
\toprule
&   &  \multicolumn{2}{c}{MAE $\downarrow$}                                & \multicolumn{2}{c}{RMSE 	$\downarrow$}                  & \multicolumn{2}{c}{$R^2$ Score 	$\uparrow$} \\ 
 \cmidrule(lr){3-4} \cmidrule(lr){5-6} \cmidrule(lr){7-8}
&  & \multicolumn{1}{c}{ID} & \multicolumn{1}{c}{OOD} & \multicolumn{1}{c}{ID} & \multicolumn{1}{c}{OOD} & \multicolumn{1}{c}{ID} & OOD     \\ 
\midrule
& One-hot    & 0.5905                 & 0.8244                 & 0.8498                  & 1.1142                 & 0.7255                           & 0.4763            \\ 
 & CGCNN      & 0.4937                 & 0.5565                 & \textbf{0.7285}         & 0.7622                 & \textbf{0.7988}                  & 0.7733            \\
 & Matscholar & 0.5183                 & 0.5635                 & 0.7714                  & 0.8058                 & 0.7737                           & 0.7448            \\
\multirow{-4}{*}{\cellcolor[HTML]{FFFFFF}Matminer Dielectric}   & MEGNet     & \textbf{0.4923}        & \textbf{0.5444}        & 0.7372                  & \textbf{0.7214}        & 0.7938                           & \textbf{0.7956}   \\
\midrule
& One-hot    & \textbf{1.1865}        & 1.4787                 & 2.6129                  & 5.4696                 & -0.0927                          & -250.7328         \\ 
 & CGCNN      & 1.3073                 & 0.8563                 & 2.5932                  & 1.1040                 & -0.0742                          & -1.7518           \\
 & Matscholar & 1.2665                 & 0.7218                 & 3.1858                  & 0.8512                 & -0.9151                          & \textbf{-0.4498}  \\
\multirow{-4}{*}{\cellcolor[HTML]{FFFFFF}Matminer Elasticity}    & MEGNet     & 1.1923                 & \textbf{0.6870}        & \textbf{2.4980}          & \textbf{0.8510}        & \textbf{0.0038}                  & -0.4580           \\

\midrule
& One-hot    & 0.5616                 & 0.9006                 & 2.7488                  & 2.2688                 & -0.2342                          & -0.3753           \\ 
 & CGCNN      & \textbf{0.5517}        & 0.6982                 & \textbf{2.6941}         & 1.8024                 & \textbf{-0.1913}                 & 0.1308            \\
 & Matscholar & 0.6619                 & \textbf{0.6397}        & 2.9217                  & \textbf{1.7527}        & -0.3963                          & \textbf{0.1797}   \\
\multirow{-4}{*}{\cellcolor[HTML]{FFFFFF}Dielectric}  & MEGNet     & 0.5850                  & 0.6810                 & 2.8342                  & 1.9833                 & -0.3102                          & -0.0524           \\

\midrule
& One-hot    & 0.0876                 & 0.2538                 & 0.1219                  & 0.3210                 & 0.9456                           & 0.7334            \\ 
 & CGCNN      & 0.0840                 & 0.1302                 & 0.1177                  & 0.1624                 & 0.9493                           & 0.9321            \\
 & Matscholar & \textbf{0.0794}        & 0.0945                 & \textbf{0.1129}         & 0.1203                 & \textbf{0.9533}                  & 0.9615            \\
\multirow{-4}{*}{\cellcolor[HTML]{FFFFFF}Perovskites}     & MEGNet     & 0.0812                 & \textbf{0.0882}        & 0.1132                  & \textbf{0.1138}        & 0.9531                           & \textbf{0.9669}   \\

\midrule
& One-hot    & 0.1424                 & 0.1843                 & 0.2035                  & 0.2359                 & 0.6923                           & 0.1683            \\ 
 & CGCNN      & 0.1310                 & 0.1190                 & 0.1935                  & 0.1545                 & \cellcolor[HTML]{FFFFFF}0.7217   & 0.6486            \\
 & Matscholar & 0.1300                 & 0.1455                 & \textbf{0.1906}         & 0.1861                 & \textbf{0.7300}                    & 0.4847            \\
\multirow{-4}{*}{\cellcolor[HTML]{FFFFFF}GVRH}     & MEGNet     & \textbf{0.1286}        & \textbf{0.1064}        & 0.1953                  & \textbf{0.1401}        & 0.7159                           & \textbf{0.7113}   \\
\midrule
& One-hot    & 0.1139                 & 0.2148                 & 0.2092                  & 0.2723                 & 0.7025                           & 0.0005            \\ 
 & CGCNN      & 0.1072                 & 0.1143                 & 0.2059                  & 0.1664                 & 0.7119                           & 0.6407            \\
 & Matscholar & \textbf{0.1036}        & 0.1149                 & \textbf{0.1926}         & 0.1706                 & \textbf{0.7481}                  & 0.6235            \\
\multirow{-4}{*}{\cellcolor[HTML]{FFFFFF}KVRH}  & MEGNet     & 0.1063                 & \textbf{0.1004}        & 0.2053                  & \textbf{0.1638}        & 0.7137                           & \textbf{0.6517}   \\
\bottomrule
\end{tabular}
\end{small}
\end{center}
\end{table}

In Table \ref{tab:er2}, we show the results of Roost trained on datasets selected using the ER2 method, where the percentages of compositions containing Ca are reduced. The performance of models on ID sets in Table \ref{tab:er2} shows a little change compared to Table \ref{tab:er1}, with physical encoding methods continuing to outperform on most datasets. However, for the Matminer Elasticity and Dielectric datasets, negative $R^2$ scores indicate that models fail to capture their properties-related information for both ID and OOD sets. For the other datasets, comparing the performance between ID and OOD sets, we observe that models using one-hot encoding deteriorate, while those using physical encoding methods improve, indicating that ER2 selection method is indeed out-of-distribution for one-hot encoding models. However, for models using physical encoding methods, the additional physical information provided in the input helps capture the correct latent composition information, making the OOD sets selected by the ER2 method less challenging than for models using one-hot encoding. In details, on the Matminer Dielectric dataset, all models show performance degradation from ID to OOD sets, except for the Matscholar encoding model, due to the small size of the training data limiting the models' extrapolation ability. The performance of the model using Matscholar encodingon the OOD set changes by -2.42\%, 1.09\%, and -1.62\% for MAE, RMSE, and $R^2$ score, respectively, indicating normal fluctuations and minimal actual performance change. On the Perovskites dataset, the model with one-hot encoding has the worst performance on both ID and OOD sets. Models using one-hot-like encoding methods (one-hot encoding and CGCNN encoding) experience more significant performance degradation than those using Matscholar encoding and MEGNet encoding. The one-hot encoding model degrades by 189.73\%, 163.33\%, and 22.44\% from ID to OOD sets, showing that even with additional Ca composition in the training data, one-hot encoding method fails to help models learn this information due to its sparsity and the limited physical information it contains. The model using CGCNN encoding, the one-hot-like encoding method, performs between than one-hot encoding but still experience significant degradation of  55.00\%, 37.98\%, and 1.81\% from the ID set to the OOD set. Models using Matscholar and MEGNet achieve between performance on both ID and OOD sets, with minimal differences in $R^2$ scores of 0.86\% and 1.45\%, respectively. On the GVRH and KVRH datasets, the one-hot encoding model continues to perform the worst. Overall, models using one-hot-like encoding methods do not perform as well as those using Matscholar and MEGNet encodings, with MEGNet encoding yielding the best results.

\begin{table}[!htb]
\caption{Results of composition-based datasets: ER2 selection method.}
\label{tab:er2}
\begin{center}
\begin{small}
\begin{tabular}{ccrrrrrr}
\toprule
&   &  \multicolumn{2}{c}{MAE $\downarrow$}                                & \multicolumn{2}{c}{RMSE 	$\downarrow$}                  & \multicolumn{2}{c}{$R^2$ Score 	$\uparrow$} \\ 
 \cmidrule(lr){3-4} \cmidrule(lr){5-6} \cmidrule(lr){7-8}
&  & \multicolumn{1}{c}{ID} & \multicolumn{1}{c}{OOD} & \multicolumn{1}{c}{ID} & \multicolumn{1}{c}{OOD} & \multicolumn{1}{c}{ID} & OOD     \\ 
\midrule
\multirow{4}{*}{Matminer Dielectric}      & One-hot    & 0.5605                 & 0.9508                  & 0.7983                 & 1.3308                  & 0.7536                 & 0.3772                  \\ 
& CGCNN      & 0.5012                 & 0.6201                  & 0.7006                 & 0.8204                  & 0.8105                 & 0.7586                  \\
& Matscholar & 0.5092                 & \textbf{0.4969}         & \textbf{0.6947}        & \textbf{0.7023}         & \textbf{0.8134}        & \textbf{0.8265}         \\
& MEGNet     & \textbf{0.4989}        & 0.5577                  & 0.7054                 & 0.7270                  & 0.8073                 & 0.8145                  \\

\midrule
\multirow{4}{*}{Matminer Elasticity}     & One-hot    & 1.0543                 & 0.5936                  & 2.7477                 & 0.8776                  & -3.2948                & -0.4237                 \\
& CGCNN      & 1.0022                 & \textbf{0.4439}         & 2.4309                 & \textbf{0.7609}         & -1.9195                & \textbf{-0.0577}        \\
& Matscholar & \textbf{0.9848}        & 0.6307                  & \textbf{1.7227}        & 0.7970                  & \textbf{-0.4996}       & -0.1682                 \\
& MEGNet     & 1.0198                 & 0.6030                  & 2.0399                 & 0.7694                  & -1.0985                & -0.1068                 \\
\midrule
\multirow{4}{*}{Dielectric}          & One-hot    & 0.6533                 & 0.7435                  & \textbf{2.9131}        & 1.7055                  & -0.3030                & 0.1189                  \\ 
& CGCNN      & 0.6231                 & \textbf{0.5905}         & 2.9184                 & \textbf{1.5584}         & \textbf{-0.3012}       & \textbf{0.2694}         \\
& Matscholar & 0.7281                 & 0.6345                  & 3.1667                 & 1.7315                  & -0.5590                 & 0.0868                  \\
& MEGNet     & \textbf{0.6101}        & 0.6307                  & 2.9697                 & 1.7981                  & -0.3534                & 0.0110                  \\
\midrule
\multirow{4}{*}{Perovskites}             & One-hot    & 0.0955                 & 0.0998                  & 0.1355                 & 0.1280                  & 0.9320                 & 0.9580                  \\ 
& CGCNN      & 0.0886                 & 0.0984                  & 0.1266                 & 0.1270                  & 0.9406                 & 0.9585                  \\
 & Matscholar & 0.0876                 & 0.0751                  & 0.1258                 & \textbf{0.0997}         & 0.9413                 & \textbf{0.9744}         \\
 & MEGNet     & \textbf{0.0834}        & \textbf{0.0741}         & \textbf{0.1187}        & 0.1028                  & \textbf{0.9478}        & 0.9728                  \\
\midrule
\multirow{4}{*}{GVRH}        & One-hot    & 0.1424                 & 0.1656                  & 0.2021                 & 0.2182                  & 0.6827                 & 0.2964                  \\ 
& CGCNN      & 0.1271                 & 0.1363                  & 0.1888                 & 0.1751                  & 0.7234                 & 0.5470                  \\
& Matscholar & 0.1267                 & 0.1246                  & \textbf{0.1810}        & 0.1613                  & \textbf{0.7457}        & 0.6167                  \\
& MEGNet     & \textbf{0.1225}        & \textbf{0.1043}         & 0.1932                 & \textbf{0.1403}         & 0.7102                 & \textbf{0.7100}         \\
\midrule
\multirow{4}{*}{KVRH}                & One-hot    & 0.1106                 & 0.1283                  & 0.2007                 & 0.2116                  & 0.7024                 & 0.4185                  \\
& CGCNN      & 0.1028                 & 0.1159                  & 0.1943                 & 0.1834                  & 0.7211                 & 0.5627                  \\
 & Matscholar & \textbf{0.0991}        & 0.1119                  & \textbf{0.1823}        & 0.1766                  & \textbf{0.7544}        & 0.5952                  \\
 & MEGNet     & 0.1024                 & \textbf{0.1032}         & 0.2024                 & \textbf{0.1734}         & 0.6974                 & \textbf{0.6091}         \\

\bottomrule

\end{tabular}
\end{small}
\end{center}
\end{table}

\begin{figure}[!htb] 
    \centering
    \begin{minipage}[c]{0.48\textwidth}
        \centering
        \includegraphics[width=\textwidth]{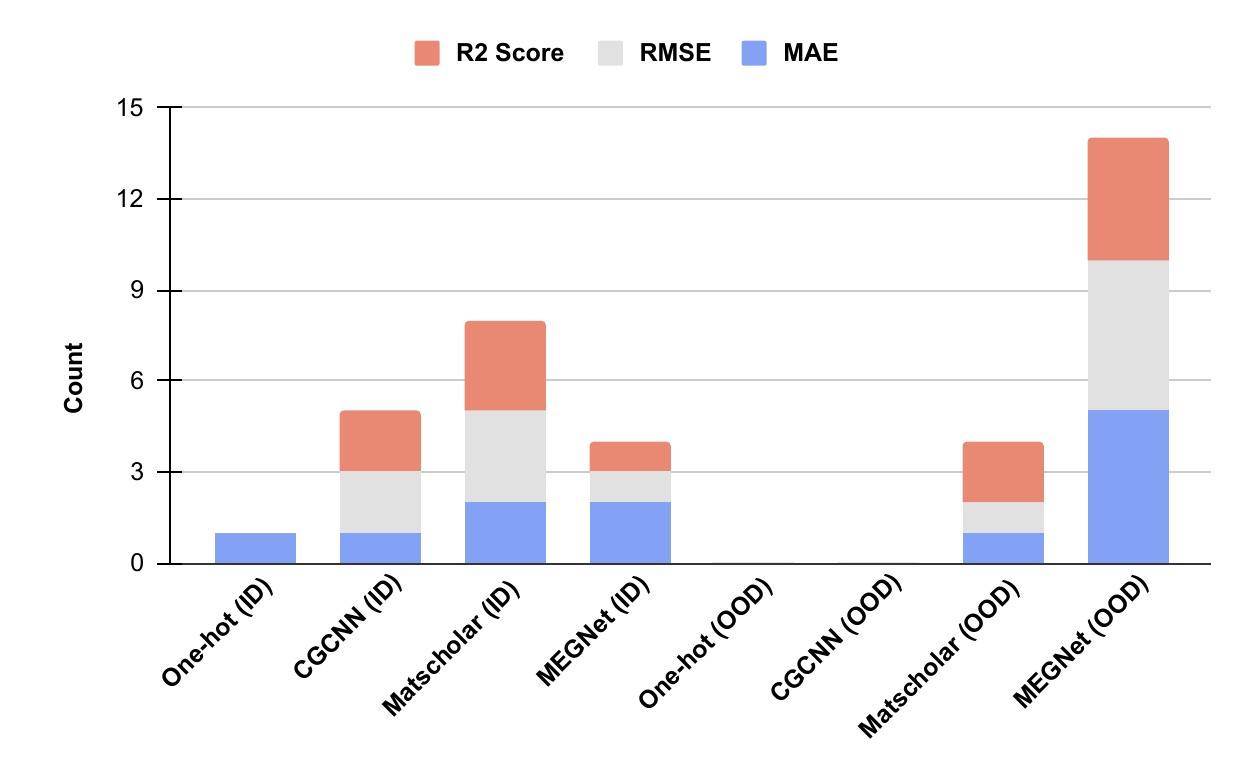}
        \subcaption{ER1 OOD selection method.}
        \label{fig:er1}
        \vspace{-1pt}       
    \end{minipage}
    \begin{minipage}[c]{0.48\textwidth}
        \centering
        \includegraphics[width=\textwidth]{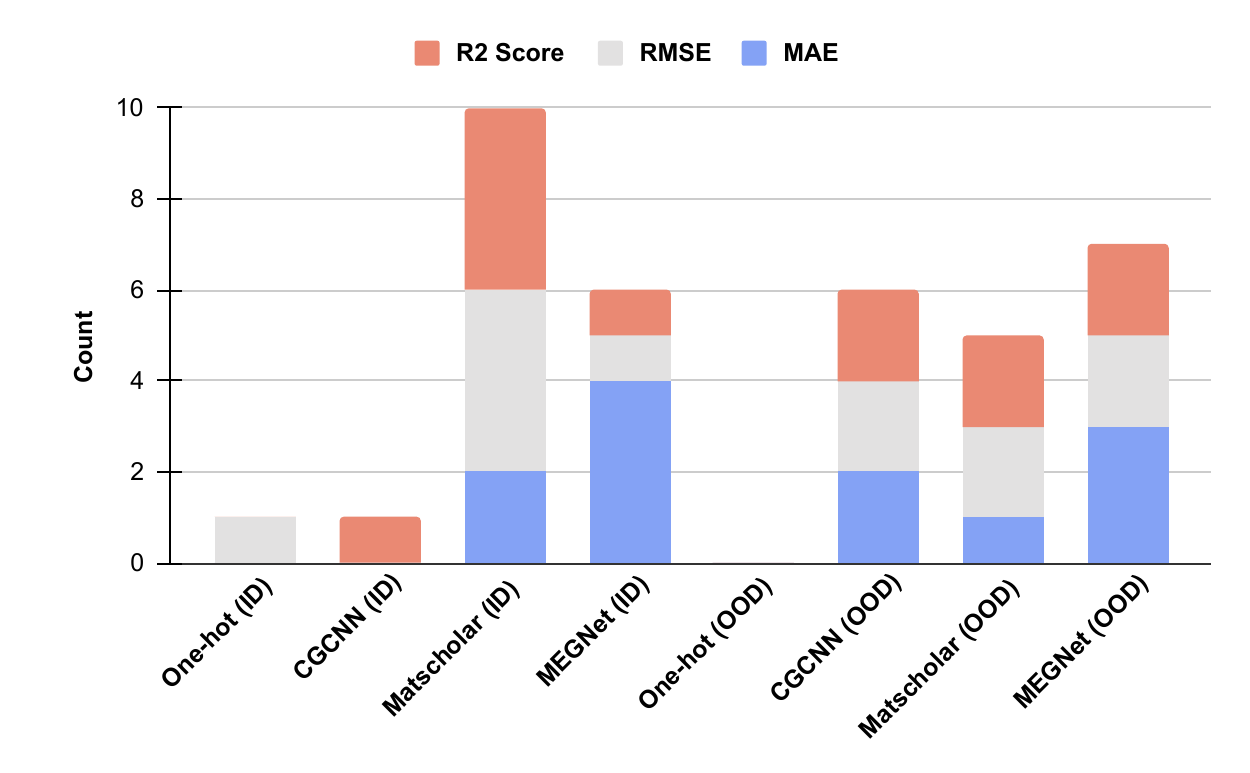}
        \subcaption{ER2 OOD selection method.}
        \label{fig:er2}
        \vspace{-1pt}       
    \end{minipage}
    \\
    \begin{minipage}[c]{0.48\textwidth}
        \centering
        \includegraphics[width=\textwidth]{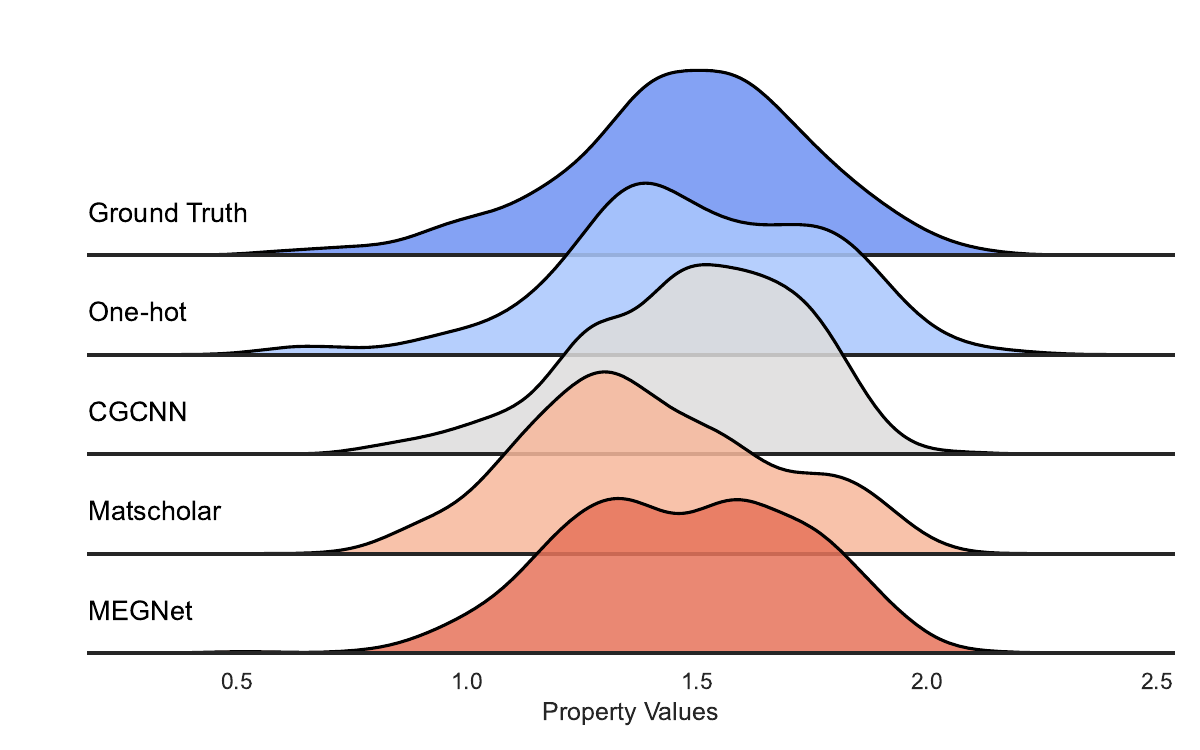}
        \subcaption{ER1 OOD selection method \& GVRH dataset.}
        \label{fig:er1gvrh}
        \vspace{-1pt}       
    \end{minipage}
    \begin{minipage}[c]{0.48\textwidth}
        \centering
        \includegraphics[width=\textwidth]{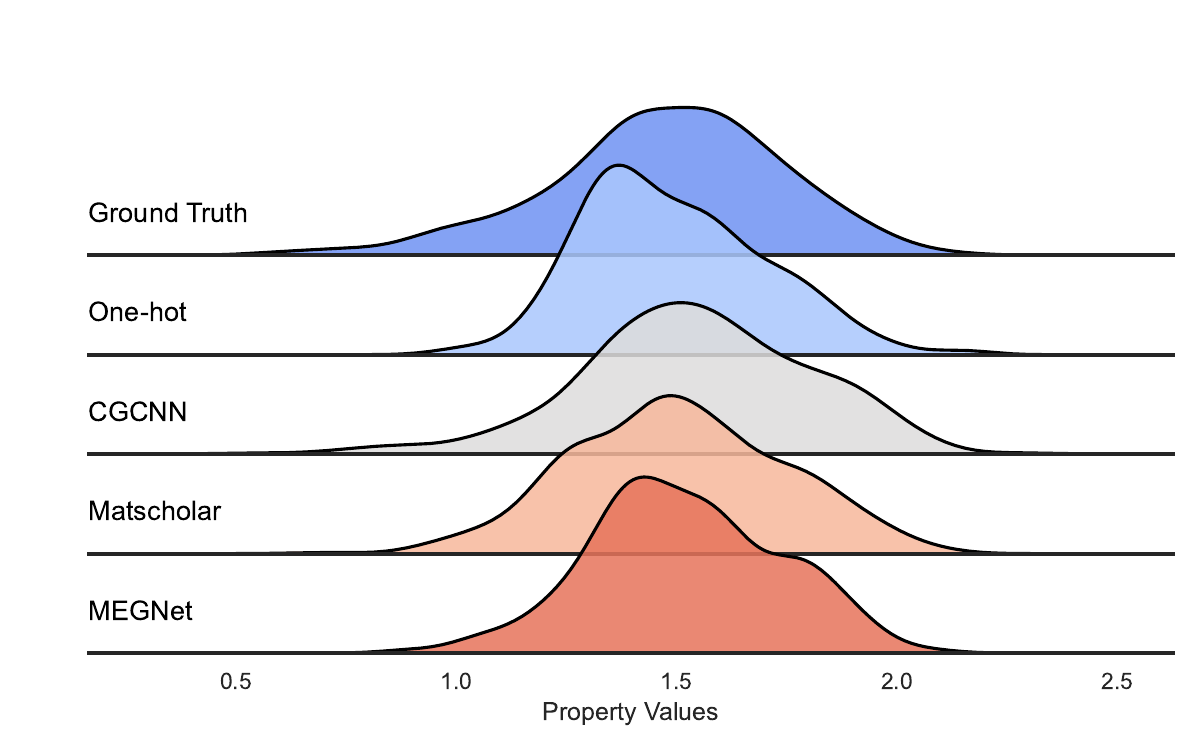}
        \subcaption{ER2 OOD selection method \& GVRH dataset.}
        \label{fig:er2gvrh}
        \vspace{-1pt}       
    \end{minipage}

     \caption{(a)-(b) Count the number of the best MAE, RMSE, and $R^2$ score for each encoding method on ID and OOD test sets: (a) ER1 OOD selection method. (b) ER2 OOD selection method; (c)-(d) Ridgeline plots for different encoding methods on GVRH datasets using ER1 and ER2 OOD selection methods, respectively. }
     \label{fig:resfig}
\end{figure}

We present a stacked column chart (Fig \ref{fig:er1} and Fid \ref{fig:er2}) to show the number of the best results (e.g., MAE, RMSE, and $R^2$ score) achieved by each encoding method. This visual clearly highlights the superior performance of physical encoding methods. In Fig \ref{fig:er1}, models using the one-hot encoding method perform poorly on both ID and OOD test sets. While CGCNN, Matscholar, and MEGNet encoding methods perform well and are similar on ID test sets, MEGNet exhibits superior performance on OOD test sets compared to the other two physical information encoding methods. This discrepancy is attributed to the ER1 selection method, where all compositions containing Ca are excluded from the training set to form the OOD test set. Consequently, the model trained with one-hot encoding never encounters Ca during training and thus cannot predict properties of compositions containing Ca in the OOD test set. For the CGCNN encoding method, although it includes more physical information by concatenating multiple one-hot encodings, the physical information is related only to the element itself, so it still struggles to predict properties of compositions it has not seen before. As a result, while CGCNN performs well on ID test sets, its performance degrades on OOD test sets. In contrast, Matscholar and MEGNet encoding methods capture not only the physical information of Ca but also its relational information. This means that even without encountering Ca directly, these models can infer relevant information. MEGNet, in particular, benefits from this approach. The latent representation from MEGNet's first layer, coupled with message passing, enables the model to capture information about elements frequently associated with Ca or similar to Ca, and so on. This additional contextual information allows MEGNet to predict the properties of compositions containing Ca (OOD test sets) despite the absence of Ca in the training data. In Fig \ref{fig:er2}, even when some compositions with Ca are included in the training sets, one-hot encoding still performs poorly on both ID and OOD test sets compared to the three physical encoding methods. However, a significant improvement is observed in the OOD performance of CGCNN compared to \ref{fig:er1}. MEGNet remains the best performance on OOD test sets, but the performance gap between the three physical encoding methods is narrower compared to \ref{fig:er1}. 

Additionally, we show the ridgeline plots for the target property (Shear Modulus) of the GVRH datasets as examples (Fig \ref{fig:er1gvrh} and Fig \ref{fig:er2gvrh}). In each figure, we compare the predicted property of models using different encoding methods with the ground truth. In Fig \ref{fig:er1gvrh}, we can find that the distribution of predicted results of the model with one-hot encoding is less like the distribution of ground truth values. For the three physical encoding methods, we can see that the model with Matscholar encoding performs worse than models using CGCNN encoding and MEGNet encoding. These results show consistency with Table \ref{tab:er1}. In Fig \ref{fig:er2gvrh}, after adding some composition containing Ca to the training data, the prediction distribution of the model using one-hot encoding is closer to the ground truth values, which is improved compared with Fig \ref{fig:er1gvrh}. However, the prediction is far less effective than the other three models. 
.

\subsection{Performance Comparison of Models for Structure-based materials property prediction}

Table \ref{tab:struct} compares the prediction performance of ALIGNN trained on three different datasets with properties selected using the FF selection method and different atomic encoding methods. On the GVRH dataset, the model using the one-hot encoding method performs worse than the other three physical encoding methods on the ID test set. Compared to the best model using the MEGNet encoding method, the MAE, RMSE, and $R^2$ score of the model using the one-hot encoding method are worse by 12.66\%, 9.38\%, and 3.21\%, respectively. For the Perovskites, the model with the one-hot encoding does not perform as poorly on both ID and OOD test sets, likely due to the simple element composition of the Perovskites dataset. On the Dielectric dataset, there is no significant difference between the model with the one-hot encoding and models using the other three physical encoding methods on the ID test set. However, on the OOD test set, the model with one-hot encoding shows a notable performance degradation, with MAE, RMSE, and $R^2$ score worse by 42.77\%, 39.84\%, and 39.84\%, respectively, compared to its own performance on the ID test set and by 7.14\%, 6.38\%, and 56.08\%, respectively, compared to the best-performing model. 

Overall, a similar trend is observed in the structure-based model: one-hot encoding struggles to capture true atomic information, leading to significant performance degradation on OOD sets compared to other physical encoding methods, especially for small training datasets.

\begin{table}[!htb]
\caption{Results of structure-based datasets: fingerprint selection method.}
\label{tab:struct}
\begin{center}
\begin{small}
\begin{tabular}{ccrrrrrr}
\toprule
&   &  \multicolumn{2}{c}{MAE $\downarrow$}                                & \multicolumn{2}{c}{RMSE 	$\downarrow$}                  & \multicolumn{2}{c}{$R^2$ Score 	$\uparrow$} \\ 
 \cmidrule(lr){3-4} \cmidrule(lr){5-6} \cmidrule(lr){7-8}
&  & \multicolumn{1}{c}{ID} & \multicolumn{1}{c}{OOD} & \multicolumn{1}{c}{ID} & \multicolumn{1}{c}{OOD} & \multicolumn{1}{c}{ID} & OOD     \\ 
\midrule
 & One-hot & 0.0979 & 0.1175 & 0.1461 & 0.1696 & 0.8433 & 0.8368         \\
& CGCNN & 0.0882 & \textbf{0.1109} & 0.1356 & 0.1641 & 0.8652 & 0.8473         \\
 & MatScholar & 0.0898 & 0.1126 & 0.1357 & \textbf{0.1623} & 0.8649 & \textbf{0.8507}  \\
\multirow{-4}{*}{\cellcolor[HTML]{FFFFFF}GVRH}   & MEGNet & \textbf{0.0869} & 0.1118 & \textbf{0.1324} & 0.1625 & \textbf{0.8713} & 0.8505     \\
 
\midrule
 & One-hot & 0.0581 & 0.0621 & 0.0870 & 0.0825 & 0.9852 & 0.9669    \\ 
 & CGCNN & 0.0561 & 0.0681 & 0.0860 & 0.0939 & 0.9855 & 0.9571   \\
 & MatScholar & \textbf{0.0548} & \textbf{0.0601} & \textbf{0.0841} & \textbf{0.0797} & \textbf{0.9861} & \textbf{0.9691}   \\
\multirow{-4}{*}{\cellcolor[HTML]{FFFFFF}Perovskites}    & MEGNet     & 0.0615 & 0.0662 & 0.0911 & 0.0868 & 0.9838 & 0.9633   \\  
\midrule
& One-hot & 0.3743 & 0.5344 & 1.5836 & 2.2145 & 0.1393 & 0.0838           \\ 
&   CGCNN & 0.4088 & 0.5039 & 1.9497 & \textbf{2.0816} & -0.5929 & \textbf{0.1908}            \\
& MatScholar & \textbf{0.3601} & 0.5139 & \textbf{1.4464} & 2.1409 & \textbf{0.2877} & 0.1441            \\
\multirow{-4}{*}{\cellcolor[HTML]{FFFFFF}Dielectric}   & MEGNet & 0.3936 & \textbf{0.4988} & 1.6643 & 2.0846 & 0.0436 & 0.1886  \\
\bottomrule

\end{tabular}
\end{small}
\end{center}
\end{table}

\section{Discussion}
\begin{figure}[ht!] 
    \begin{subfigure}[t]{0.33\textwidth}
        \includegraphics[width=0.99\textwidth]{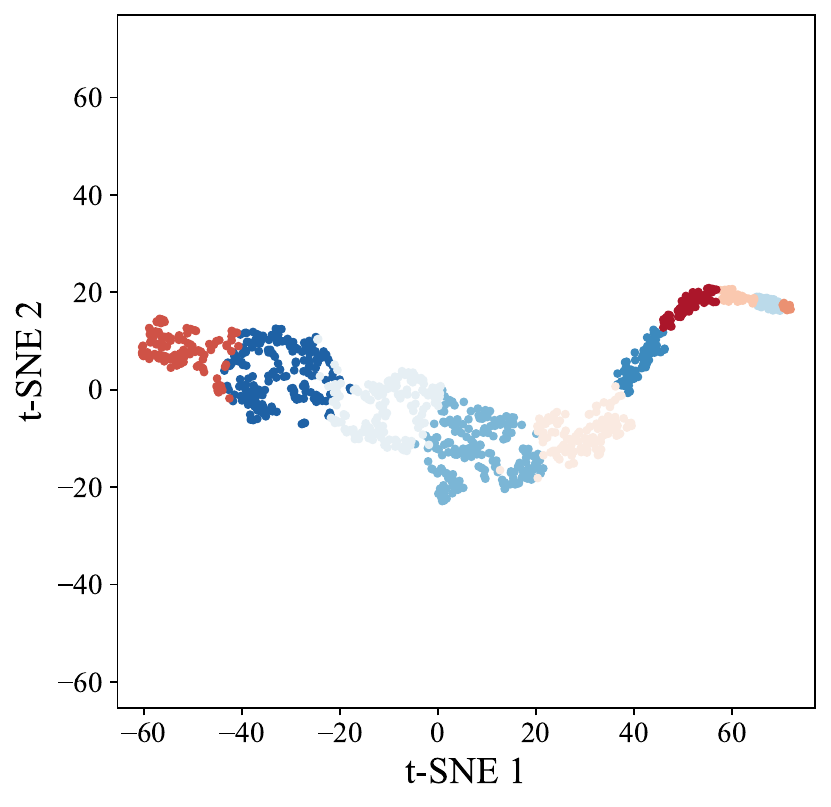}
        \caption{Matminer Dielectric}
        \vspace{3pt}
        \label{fig:matminer_dielect}
    \end{subfigure}
 \begin{subfigure}[t]{0.33\textwidth}
        \includegraphics[width=0.99\textwidth]{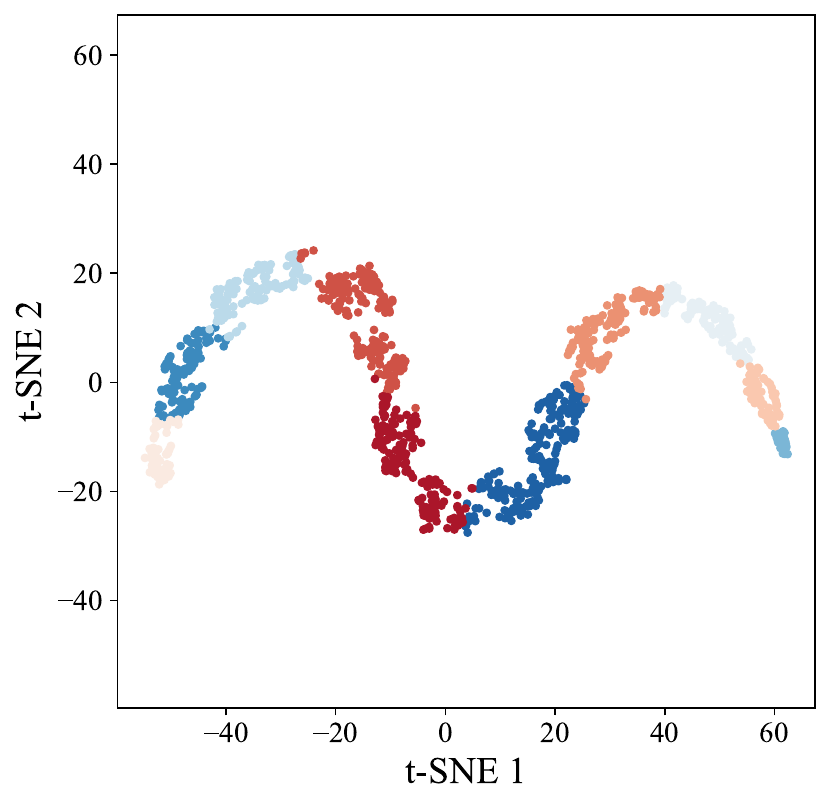}
        \caption{Matminer Elasticity}
        \vspace{-3pt}
        \label{fig:matminer_elast}
    \end{subfigure}    
    \begin{subfigure}[t]{0.33\textwidth}
        \includegraphics[width=0.99\textwidth]{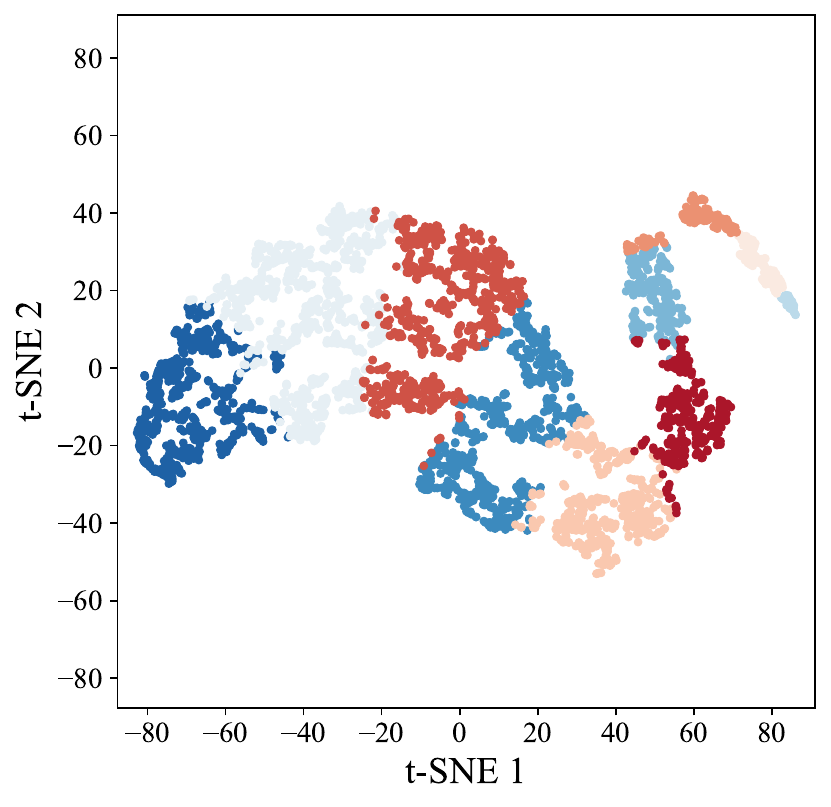}
        \caption{Dielectric}
        \vspace{3pt}
        \label{fig:dielect}
    \end{subfigure} 
    \begin{subfigure}[t]{0.33\textwidth}
        \includegraphics[width=0.99\textwidth]{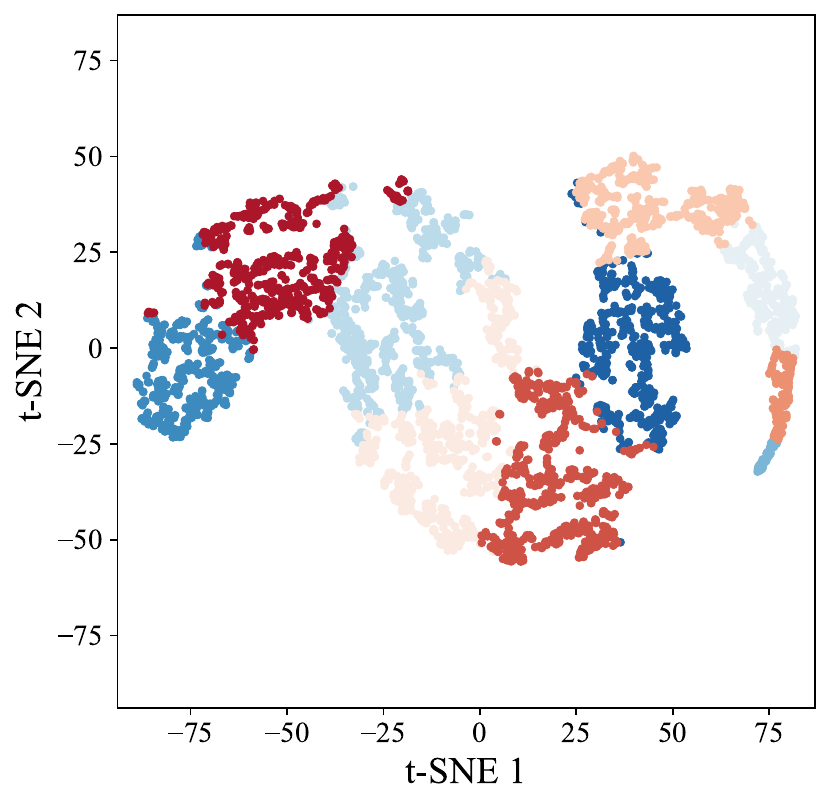}
        \caption{Perovskites}
        \vspace{-3pt}
        \label{fig:perov}
    \end{subfigure}
    \begin{subfigure}[t]{0.33\textwidth}
        \includegraphics[width=0.99\textwidth]{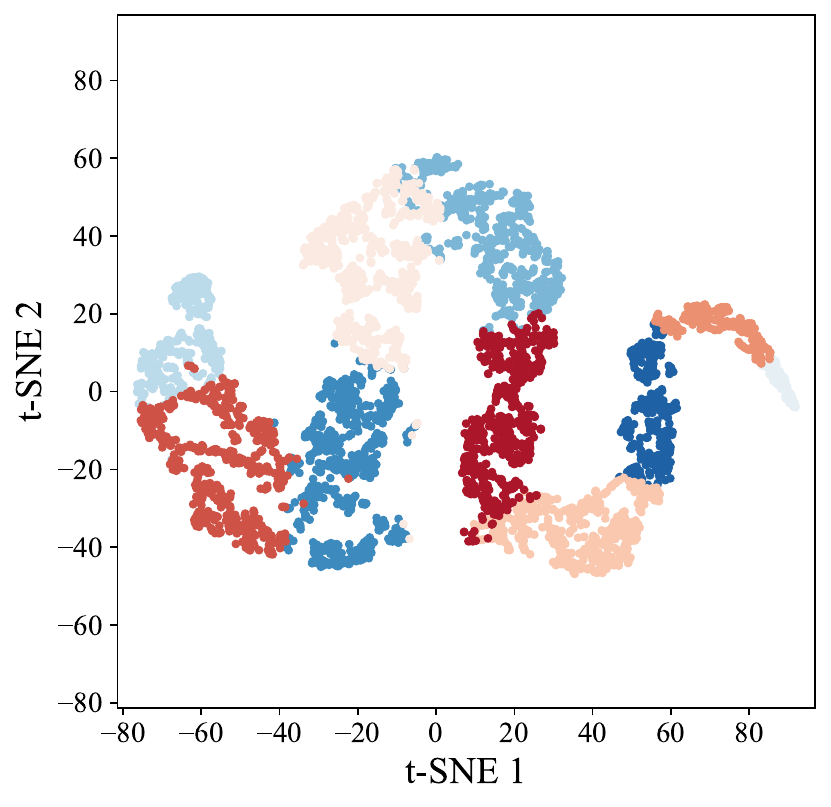}
        \caption{GVRH}
        \vspace{-3pt}
        \label{fig:gvrh}
    \end{subfigure} 

  \caption{t-SNE plots for different datasets using cluster OOD selection method. We use 10 colors to represent 10 different clusters for each dataset. During the training process, we selected one cluster as the OOD test set and the other clusters as the training datasets.}
  \label{fig:cluster}
\end{figure}

A key issue in OOD performance evaluation is how to select the OOD test set. 
During our experiments, we also explored another different OOD selection method (cluster OOD selection method) based on the Magpie features of material compositions as recommended in \cite{meredig2018can}. Magpie features \cite{ward2016general} are derived from the elemental composition of materials and represent their properties. In the cluster OOD selection method, we applied K-means clustering to divide each dataset into 10 clusters as illustrated in Fig \ref{fig:cluster}. During the training process, we selected each cluster as the OOD test set while using the other clusters as the training set, resulting in ten different trained models. We then averaged their performance to obtain the final results. Note that GVRH and KVRH share the same compositions, so we only display one cluster figure in Fig \ref{fig:cluster}. 

Logically, the clusters should differ significantly from one another, and Fig \ref{fig:cluster} also showed that distinct areas are occupied by different areas, suggesting that each cluster is mutually out-of-distribution relative to the others. However, our quantitative results of the Roost models revealed that the models' performance on the OOD test sets was often better than those on the ID test sets. This was consistently observed across most models. This outcome led us to reflect and conclude that selecting an OOD set based on input characteristics (e.g., using Magpie features) does not necessarily guarantee an OOD distribution in the output layer, especially for deep learning models. Consequently, we adopted different OOD selection methods in our study.

\section{Conclusion}

This work evaluates the performance of deep learning models trained with various node encoding methods on both in-distribution (ID) and out-of-distribution (OOD) test sets across six different datasets. The encoding methods investigated include one-hot encoding, CGCNN encoding, Matscholar encoding, and MEGNet encoding, each representing different approaches to capturing elemental information. Our experimental results indicate that while one-hot encoding achieves reasonable performance on ID test sets, it consistently underperforms on OOD test sets. This suggests that one-hot encoding fails to generalize effectively to new and unseen data due to its simple and sparse nature, which does not adequately capture the underlying physical and chemical properties of the elements, especially when the training set is small. 

These findings of this paper underscore the importance of selecting appropriate encoding methods for node features in DL models, particularly when dealing with OOD scenarios. Encoding methods that capture more comprehensive information about the elements significantly enhance the models' ability to learn meaningful patterns and make accurate predictions on new materials. Future research may continue to explore and refine the encoding techniques, focusing on enhancing the representation of elemental properties to further improve the generalization of DL models in materials science.

\section{Data Availability}
Details about the datasets can be found at the links in the references listed in Table 1.

\section*{Contributions}
Conceptualization, J.H.; methodology, N.F., S.O., and J.H.; software, N.F. and S.O.; writing--original draft preparation, N.F., S.O., and J.H.; writing--review and editing, N.F., S.O., and J.H.; visualization, N.F. and S.O.; supervision, J.H.

\section*{Acknowledgments}
The research reported in this work was supported in part by National Science Foundation under the grant 2110033, OAC-2311202, and 2320292. The views, perspectives, and content do not necessarily represent the official views of the NSF.

\bibliography{references}
\bibliographystyle{unsrt}

\newpage
\appendix

\end{document}

%% file: math_commands.tex
\usepackage{amsmath,amsfonts,bm}

\def\eqref#1{equation~\ref{#1}}

\def\1{\bm{1}}

\DeclareMathAlphabet{\mathsfit}{\encodingdefault}{\sfdefault}{m}{sl}
\SetMathAlphabet{\mathsfit}{bold}{\encodingdefault}{\sfdefault}{bx}{n}